\documentclass[twocolumn,showpacs,showkeys,preprintnumbers,amsmath,amssymb]{revtex4}

\usepackage{graphicx}
\usepackage{dcolumn}
\usepackage{bm}

\begin{document}

\preprint{APS/123-QED}

\title{$\Delta v=2$ seniority changing transitions in yrast $3^-$ states and B(E3) systematics of Sn isotopes}
\author{Bhoomika Maheshwari}
\email{bhoomika.physics@gmail.com}
\affiliation{Department of Physics, Banasthali University, Banasthali-304022, India.}
\author{Swati Garg}
\author{Ashok Kumar Jain}%
\affiliation{Department of Physics, Indian Institute of Technology, Roorkee 247667, India.}
\date{\today}

\begin{abstract}
We show for the first time that the generalized seniority scheme explains reasonably well the $B(E3)$ systematics for the $(0^+ \rightarrow 3_1^-)$ transitions in the Sn-isotopes, which are odd-tensor $E3$ transitions connecting different seniority states ($\Delta v = 2$). Additionally, we also present Large Scale Shell Model (LSSM) calculations to support our interpretation. The generalized seniority scheme points to the octupole character of these $3^-$ states in Sn isotopes.”
\end{abstract}

\maketitle

Symmetries in physics play a fundamental role in the theoretical description of a wide range of phenomena and are particularly useful in systematizing the properties and underlying structure of nuclei. The concept of seniority has proved to be a powerful tool in exploring nuclei close to the magic numbers, and may also be related to the symmetry in pairing of nucleons. The seniority scheme was first introduced by Racah~\cite{racah} in the atomic context to distinguish the states having same values of $L$, $S$, and $J$ in $LS$-coupling, where $L$, $S$, and $J$ represent the orbital, spin and total angular momentum, respectively. In 1950s, two independent groups viz. Racah and Talmi~\cite{Racah1952}, and Flowers~\cite{Flowers1952} introduced the concept of seniority in nuclear physics. The usefulness of seniority has continued to grow since then. Seniority may simply be defined as the number of unpaired nucleons for a given state, and is generally denoted by $v$. This means that the state with an odd-particle in the valence orbit $J=j$, always has the lowest seniority as $v=1$. On the other hand, the ground states of even-even nuclei always have the lowest seniority as $v=0$, i.e. all the particles are paired to $J=0$ state. It is now well known that $v$ remains a good quantum number for states emanating from a pure-j configuration with j$\le$7/2, as there is no chance of seniority mixing. This description becomes cumbersome for higher-j values. However, the validity of seniority for higher-j values has also been suggested and tested depending upon the nucleon configuration and states. A generalization of the single-j pure seniority scheme to multi-j generalized seniority was first proposed by Arima and Ichimura~\cite{arima}. Generalized seniority is able to take care of the presence of multi-j orbits in a given nuclear state. 

The seniority scheme has been quite successful in explaining the spectroscopic features of semi-magic nuclei. Sn-isotopes represent the longest chain of isotopes in the nuclear chart. Because of the semi-magic nature of the Sn-isotopes, the seniority scheme ~\cite{racah, talmi, casten} has been extensively used to understand the various systematic features in these nuclei ~\cite{morales, astier}. They also provide a rigorous testing ground for the various effective interactions used in the large scale shell model (LSSM) calculations ~\cite{simpson, maheshwari}. Extensive calculations by using the generalized seniority scheme and the shell model for N=82 isotopes were reported by Sagawa et al.~\cite{sagawa}, where they calculated the quadrupole transition rates while addressing the problem of core polarization and effective charges for protons and neutrons. We have recently used the generalized seniority formalism for multi-j degenerate orbits to show ~\cite{maheshwari1} that both the even and odd tensor electric transition probabilities for same seniority ($\Delta v = 0$) transitions follow similar trends in high-spin isomers ~\cite{atlas}, as schematically shown by the solid line in Fig.~\ref{fig:bel}. Here, the reduced transition probabilities, $B(EL)$, follow a parabolic behavior with a minimum at the middle of the valence space. As a result, one expects larger half-life seniority isomers in the middle of the shell for both the even and the odd electric transitions. In our previous work ~\cite{maheshwari1}, we have convincingly shown for the first time that the odd tensor $E1$, $\Delta v = 0$ transition probabilities from the ${13}^-$ isomers in Sn-isotopes behave similar to the even tensor $E2$, $\Delta v = 0$ transition probabilities of the ${10}^+$ and ${15}^-$ isomers in Sn-isotopes. As a consequence, the two classes of isomers may be treated on the same footing in this simple approach. This immediately expands the scope of seniority isomers to odd tensor electric transitions also. The generalized seniority scheme has also been used by us to interpret the various high spin isomers in the $Z=50$, $N=82$ and $Z=82$ chains and compare their similarities and differences~\cite{china}. This scheme also provides an insight in the location of single-particle energies and the effective interactions in neutron-rich nuclei~\cite{scripta}. To sum up, the generalized seniority scheme provides an unique probe to explore the semi-magic nuclei and their immediate neighbors. 

It was also pointed out by us that the $B(EL)$ values for either odd or even $L$ in the case of seniority change by $\Delta v = 2$ exhibit an inverted parabolic behavior, as shown by the dashed line in Fig.~\ref{fig:bel}. We have recently used the seniority changing $\Delta v = 2$ even tensor transitions in the case of $B(E2 \uparrow)$ values for the $0^+$ to $2^+$ transitions in the Sn-isotopes to explain the two asymmetric parabolas~\cite{maheshwari2}. We are aware of the fact that many independent groups are involved in studying the first excited $2^+$ states in Sn isotopes. However, a very limited amount of work  has been done for the first excited $3^-$ states in the Sn isotopes. In this paper, we focus our attention on the $3^-$ states and show that the $B(E3 \uparrow)$ values for the $0^+$ to $3^-$ transitions in the Sn-isotopes nicely fit into the inverted parabolic behavior from the calculations obtained by our scheme. This is the first evidence of the seniority changing $\Delta v = 2$ odd tensor $E3$ transitions confirming the results of the generalized seniority scheme presented by us.

To show this, we have used the $B(E3 \uparrow;0^+ \rightarrow 3_1^-)$ values of the first $3^-$ states in the Sn-isotopes from the compilation of Kibedi and Spear ~\cite{kibedi}, which decay to the $0^+$ ground state via $E3$ transition. These $3^-$ states are generally believed to be octupole vibrational in character. In 1981, Jonsson $et$ $al.$~\cite{jonsson} reported the measurements of $3^-$ states in $^{112-124}$Sn using $(p,p' \gamma)$ reaction and Coulomb excitation, and also compared their measurements with previously known data and some theoretical calculations which were quite far from the experimental data. The behavior of the $3^-$ states was later studied by Ansari and Ring ~\cite{ansari} within the relativistic quasiparticle random-phase approximation (RQRPA), which highlighted the new challenges in fixing the force parameters for the $NL1$ and $NL3$ interactions to reproduce the experimental data. The RQRPA estimates were successful in obtaining the overall trend of the $B(E3)$ values although the actual values differed quite a bit from theory. We now present a study of the $3^-$ states by using the simple microscopic approach based on generalized seniority scheme reported in our earlier paper ~\cite{maheshwari1}. We find that the same scheme reproduces the systematics of $B(E3 \uparrow,0^+ \rightarrow 3_1^-)$ transition probabilities quite well. This work reports the understanding of these states as the generalized seniority states for the first time. The involvement of $d$ and $h$ orbits is found to be essential to explain the nature of the $B(E3)$ values, which obviously fits into the octupole nature of the $3^-$ states. This understanding supports the previous interpretation of these states being octupole vibrational in character, as the $d$ and $h$ orbits can be connected by a $\Delta l=3$ interaction only. 

We have also carried out the LSSM calculations to validate our results. These LSSM calculations are, however, restricted in nature due to our computational limitations. Even then, they serve the basic purpose by supporting the interpretation that emerges from the generalized seniority approach. The generalized seniority scheme also turns out to be immensely useful in singling out the most important orbitals involved in the LSSM calculations. We present the details of the calculations and results in the next section. This simple approach may also prove useful for estimating the transition probabilities in those cases for which measurements have not been made so far. 

\section{Calculations and results}

We have used the generalized seniority formalism for multi-j degenerate orbits presented in our recent paper~\cite{maheshwari1}, to calculate the $B(E3)$ values in the Sn-isotopes. The reduced $B(E3)$ transition probabilities for $n$ particles in the multi-j $\tilde{j}=j \otimes j' ....$ configuration can be obtained from the equation, 
\begin{eqnarray}
B(E3) \uparrow=\frac{7}{2J_i+1}|\langle \tilde{j}^n v l J_f || \sum_i r_i^3 Y^{3}(\theta_i,\phi_i) || \tilde{j}^n v' l' J_i \rangle |^2
\end{eqnarray}
where the reduced matrix elements between $\Delta v = 2$ states can be written in terms of seniority reduction formula as [5],
\begin{eqnarray}
\langle \tilde{j}^n v l J_f ||\sum_i r_i^3 Y^{3}|| \tilde{j}^n v\pm 2 l' J_i \rangle  = \nonumber\\ \Bigg[ \sqrt{\frac{(n-v+2)(2\Omega+2-n-v)}{4(\Omega+1-v)}} \Bigg] \nonumber\\ \langle \tilde{j}^v v l J_f ||\sum_i r_i^3 Y^{3}|| \tilde{j}^v v\pm 2 l' J_i \rangle 
\end{eqnarray}

The coefficients in the square brackets depend on the particle number $n$, the generalized seniority $v$ and the corresponding total pair degeneracy $\Omega= \frac{1}{2}(2 \tilde{j} +1)= \frac{1}{2} \sum \limits_j (2j+1)$ in the multi-j configuration. Note that the reduced matrix elements on the left side of the equation are in $\tilde{j}^n$ configuration, while the reduced matrix elements on the right side are in $\tilde{j}^v$ configuration. The said equation simply reduces the particle number $(n)$ dependency of these matrix elements to the generalized seniority $(v)$. This means that the calculated trend will simply depend on the coefficients in the square brackets. We have calculated the $B(E3) \uparrow$ values for the $3^-$ states in the Sn-isotopes by using two values of $\Omega= 9$ and $11$, corresponding to the $d_{5/2} \otimes h_{11/2}$, and $d_{5/2} \otimes d_{3/2} \otimes h_{11/2}$ valence spaces, respectively. It is quite obvious that the mixing of negative parity $h_{11/2}$ orbital is required to generate $3^-$ states, which decay via $E3$ transitions to ground state. This points towards the octupole character of the transition and hence, the incorporation of the d-orbital is necessary. 

\begin{figure}
\includegraphics[width=9.5cm,height=9cm]{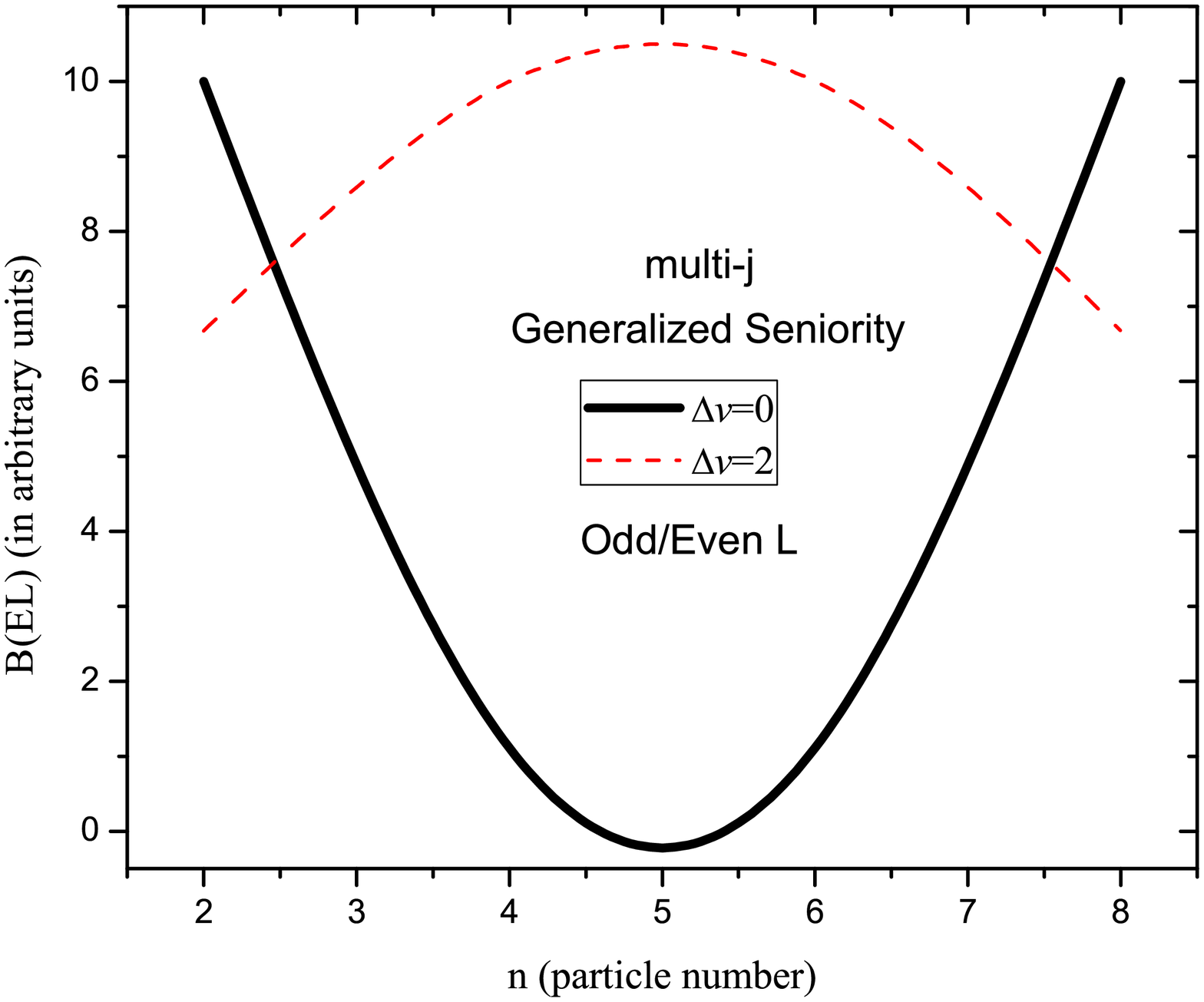} 
\caption{\label{fig:bel}(Color online) Schematic plot of the $B(EL)$ values, for both even and odd $L$, with particle number $n$, for seniority conserving $\Delta v=0$ transitions (solid line) and seniority changing $\Delta v=2$ transitions (dashed line), using a pair degeneracy of $\Omega=5$.} 
\end{figure}

\begin{figure}
\includegraphics[width=9.5cm,height=9cm]{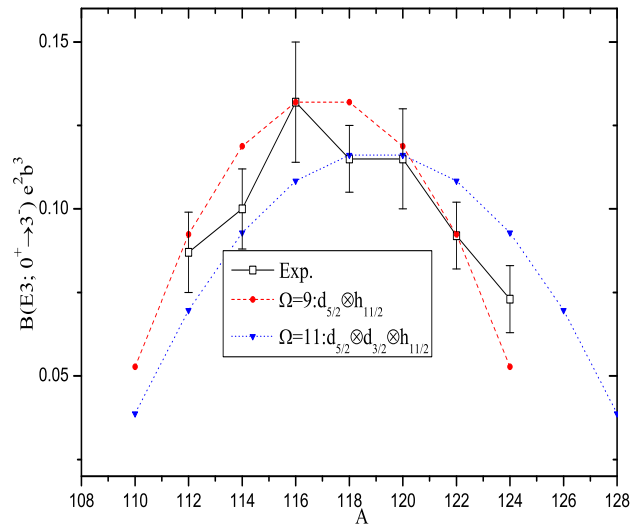}
\caption{\label{fig:be3}(Color online) A comparison of the experimental $B(E3)$ systematics in the Sn-isotopes (solid line) with the generalized seniority calculations. The dashed and dotted lines stand for a choice of the $d_{5/2} \otimes h_{11/2}$, and $d_{5/2} \otimes d_{3/2} \otimes h_{11/2}$ valence spaces, respectively.} 
\end{figure}

We, therefore, calculate the complete systematics by fitting the experimental $B(E3)$ values for $^{116}$Sn and $^{118}$Sn, the respective mid-points, for $\Omega=9$ and $11$. We have taken $^{108}$Sn as the core, where the core represents the $n=0$ situation and represents the situation corresponding to complete filling of the $g_{7/2}$ orbit. The calculations have been done by assuming $\Delta v =2$ transitions in the generalized seniority scheme. A comparison of the calculated results for both the $\Omega$ values with the known experimental data~\cite{kibedi} is shown in Fig.~\ref{fig:be3}. The solid line stands for the experimental data while the dashed and dotted lines stand for the generalized seniority calculations by using $\Omega=9$ and $11$, respectively. The measured values as well as the calculations show a peak in the middle of the given valence space. This behavior is as expected for odd tensor transitions that connect the states having seniorities differing by $\Delta v=2$ (Fig.~\ref{fig:bel}). Hence, the simple generalized seniority scheme is able to explain the complete systematics with the valence space consisting of $d$ and $h$ orbits. The calculated numbers with $\Omega=9$ are in better agreement indicating the configuration for these states is pre-dominantly $d_{5/2} \otimes h_{11/2}$. The calculations with $\Omega = 11$ spoil the results, which strongly suggests that the incorporation of other orbitals does not help, and validates the configuration corresponding to $\Omega = 9$.

\begin{figure}
\includegraphics[width=9.5cm,height=9cm]{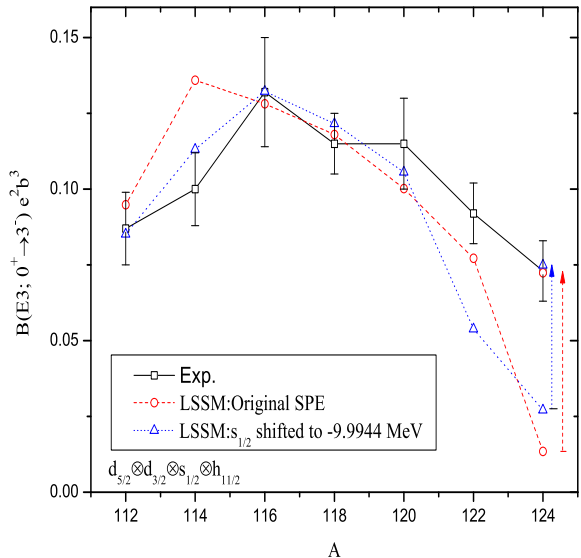}
\caption{\label{fig:BE3_full}(Color online) A comparison of the experimental $B(E3)$ trend with the LSSM results for the $d_{5/2} \otimes d_{3/2} \otimes s_{1/2} \otimes h_{11/2}$ valence space having original SPE, and $s_{1/2}$ shifted to -9.9944 MeV, respectively.} 
\end{figure}

We have also carried out the LSSM calculations by using the Nushell code~\cite{brown} and SN100PN interaction~\cite{brown1} for Sn-isotopes with original single particle energies. The neutron single particle energies have been taken as -10.6089, -10.2893, -8.7167, -8.6944, -8.8152 MeV for the available $0g_{7/2}$, $1d_{5/2}$, $1d_{3/2}$, $2s_{1/2}$, and $0h_{11/2}$ valence orbits. We have chosen the $g_{7/2}$ orbit to be completely filled. The harmonic oscillator potential was chosen with an oscillator parameter of $\hbar \omega =45A^{-1/3}-25A^{-2/3}$. The scaling factor for the two-body matrix elements of the interaction varies as $A^{-1/3}$. We have compared the calculated results along with the experimental data~\cite{ensdf} in Fig.~\ref{fig:BE3_full}. These calculations reproduce the measured values quite well, except for a peak at $^{114}$Sn in the calculated values. 

To resolve this, we shift the single-particle energy (SPE) of the $s_{1/2}$ orbit by 1.3 MeV to -9.9944 MeV, and carry out the LSSM calculations with the $d_{5/2} \otimes d_{3/2} \otimes s_{1/2} \otimes h_{11/2}$ valence space. We find that the shift in SPE of the $s_{1/2}$ orbit brings the $B(E3)$ value in agreement with the data, though the values on higher mass side get disturbed. We note that an unusually large effective charge value of $1.6$ has been used in these calculations. On the other hand, we obtain the two-body reduced transition matrix elements by fitting to one of the measured values. This fitting takes care of the dependence on the other parameters, like single particle energies, effective charges etc.

We conclude that the partial gap at $N=64$ in SPE of the orbits is responsible for a kink at $^{114}$Sn, which gets washed out by shifting the $s_{1/2}$ orbit as the gap now shifts to $N=66$. We feel that an alternate and efficient approach to renormalize may be to carry out a tuning of the two-body matrix elements for the new truncated valence spaces, which is beyond the scope of our work. We could perform a full space calculations only for $^{124}$Sn with our present computational facilities. We present the $B(E3)$ values at $^{124}$Sn from the full space calculations for both the sets of SPE, i.e. with the original set of SPE and the new set with $s_{1/2}$ orbit shifted to -9.9944 MeV; we can see that the full space calculations are able to reproduce the observed value very well as shown in Fig.~\ref{fig:BE3_full} by the triangle and the circle almost superposed on the measured value. Therefore, much of the disagreement with observed data could be due to the truncated space used in our calculations. We also note the limitation that the code is only using the two-body matrix elements of $d_{5/2}$ and $h_{11/2}$ orbits for the $B(E3)$ calculations, as these orbits have $\Delta j=3$.

However, the deviation of the shell model results for $^{114}$Sn isotope has been very puzzling. This isotope has $N=64$, a possible magic number for octupole collectivity~\cite{nazarewicz, cottle}. Yet, the measured $B(E3)$ value for $^{114}$Sn is observed to be lower than $^{116}$Sn. Also, the measured $B(E3)$ behavior exhibits a peak at $^{116}$Sn, while our LSSM calculations (with original SPE) have a sharp peak at $^{114}$Sn as shown in the Fig.~\ref{fig:BE3_full}, unless we shift the SPE of $s_{1/2}$ orbit, and change this partial gap to $N=66$. This shifting resolves the kink in the $B(E3)$ calculations and reproduces the experimental trend as shown in Fig.~\ref{fig:BE3_full}. It is interesting to note that the peak in the $B(E3)$ value corresponds to the situation where the $h_{11/2}$ and $d_{5/2}$ orbits saturate at one-particle, and one-hole configuration, respectively. The next nucleons start filling in $h_{11/2}$ orbit along with one-hole in $d_{5/2}$, as both the $d_{5/2}$ and $h_{11/2}$ orbits are necessary to generate the $3^-$ state.

The results of the RQRPA calculations of Ansari and Ring~\cite{ansari} although far from the experimental values, do show a fall in the $B(E3)$ value for $^{114}$Sn as seen in the experimental data. These calculations are noteworthy as they are able to obtain the qualitative trend reasonably well. Our calculations by using the generalized seniority scheme also reproduce the results reasonably well. On the other hand, the LSSM calculations are not able to reproduce the trend in $B(E3)$ at $^{114}$Sn without shifting the $s_{1/2}$ orbit. However, we feel that the LSSM calculations in full space are needed to arrive at a final judgment.   

\section{Conclusion}

To conclude, we have used the generalized seniority scheme for multi-j degenerate orbits to calculate the $B(E3 \uparrow,0^+ \rightarrow 3_1^-)$ transition probabilities for the $3^-$ states in the Sn-isotopes. Our calculations successfully reproduce the parabolic behavior of $B(E3)$ values with a peak in the middle and support the interpretation that the transition involves a seniority change of $2$. Our explanation also supports the octupole vibrational character of the $3^-$ states, as they mainly arise due to $d$ and $h$ orbits differing in $l$ values by $3$ units. We also present LSSM calculations to support the generalized seniority scheme; the results of LSSM calculations support our interpretation except for a puzzling deviation at $^{114}$Sn. The deviation in the shell model results at $^{114}$Sn is sought to be explained by shifting the $s_{1/2}$ orbit to create a partial gap at $N=66$ instead of $N=64$. Note that the LSSM calculations have been done by using a truncated space due to our computational limitations. Still, the shell model results serve the basic purpose of validating our interpretation. We feel that not only new measurement efforts are required to obtain more precise data but full space LSSM calculations should also be carried out.

\section*{Acknowledgments}

Financial support from the Ministry of Human Resource Development (Government of India), in the form of a doctoral fellowship to B.M., is gratefully acknowledged.

\newpage 
\bibliography{apssamp}
  
\end{document}